\documentclass[twocolumn,prl,floatfix]{revtex4-1}
\usepackage{xcolor} 
\usepackage{graphicx}
\usepackage{amssymb}
\usepackage{amsmath}
\usepackage[utf8]{inputenc}

\usepackage{tikz}
\newcommand{\op}[1]{\hat{#1}}
\newcommand{\mat}[1]{\mathrm{#1}}

\newcommand{\change}[1]{\textcolor{black}{#1}}

\begin{document}
\title{Identification of strongly interacting organic semimetals}

\author{R. Matthias Geilhufe$^{1}$}
\email{matthias.geilhufe@su.se}
\author{Bart Olsthoorn$^{1}$}
\affiliation{$^1$Nordita,  KTH Royal Institute of Technology and Stockholm University, Roslagstullsbacken 23,  10691 Stockholm,  Sweden}

\date{\today}

\begin {abstract}
Dirac- and Weyl point- and line-node semimetals are characterized by a zero band gap with simultaneously vanishing density of states. Given a sufficient interaction strength, such materials can undergo an interaction instability, e.g., into an excitonic insulator phase. Due to generically flat bands, organic crystals represent a promising materials class in this regard. We combine machine learning, density functional theory, and effective models to identify specific example materials. Without taking into account the effect of many-body interactions, we found the organic charge transfer salts (EDT-TTF-I$_2$)$_2$(DDQ)$\cdot($CH$_3$CN) and TSeF-TCNQ and a bis-1,2,3-dithiazolyl radical conductor to exhibit a semimetallic phase in our ab initio calculations. Adding the effect of strong particle-hole interactions for (EDT-TTF-I$_2$)$_2$(DDQ)$\cdot($CH$_3$CN) and TSeF-TCNQ opens an excitonic gap in the order of 60 meV and 100 meV, which is in good agreement with previous experiments on these materials. 

\end{abstract}
\maketitle

\section{Introduction}
Semimetals have attracted huge attention due to their striking transport properties, analogies to high energy physics phenomena, and potential for functionalization \cite{wehling2014dirac,vafek2014,wang2017quantum}. Their realization relies on a delicate combination of symmetry, electron-filling, and band ordering enforcing the existence of the nodes in the band structure at the chemical potential while having a vanishing density of states (DOS) at the crossing point \cite{yang2014classification,gao2016classification,watanabe2016filling}. It has been shown extensively for the case of Dirac semimetals, that under a sufficiently high interaction strength a dynamical mass term can be generated leading to a quantum phase transition into a gapped phase \cite{pisarski1984chiral, jurivcic2009coulomb, gonzalez2015phase, hofmann2015interacting}. This quantum phase transition strongly depends on: i) the effective fine structure constant $\alpha_{\text{eff}}$, describing the ratio of the coupling of the Fermion field to its gauge field versus the kinetic energy; ii) the dimension of the system; iii) the number of fermionic flavors. Similar phenomena where also discussed in the case of \change{Weyl semimetals \cite{roy2017interactingWeyl,roy2018global} and line-node semimetals \cite{sur2016instabilities,gao2016classification,roy2017interacting}}.

With the goal of identifying experimentally feasible materials to investigate interaction effects in nodal semimetals we focus on organic crystals. Organic crystals typically exhibit strong intramolecular forces and weak intermolecular forces, leading to tiny hopping amplitudes for electrons between molecules and resulting flat electronic bands. The flatness corresponds to a tiny quasiparticle kinetic energy and dominant interaction effects. We focus on the excitonic insulator state occurring when a weakly screened Coulomb interaction between a hole and an electron leads to an electron-hole bound state \cite{mott1961transition,jerome1967excitonic}.

Even though organics seem to be promising materials for strong interaction effects, we face several major difficulties: i) the search space is massive, e.g., the crystallographic open database stores $\approx 200,000$ crystal structures containing carbon and hydrogen \cite{gravzulis2009crystallography}; ii) organics are typically large band gap insulators \cite{borysov2017organic}; iii) complex unit cells and strong correlation effects are challenging for an ab initio description. Hence, we apply the following procedure: first, we apply machine learning to narrow down the search space to a computationally feasible set of materials which are predicted to show a tiny band gap; second, using density functional theory we compute the band structures for these materials; third, we select the occurring semimetals, construct effective electronic models and numerically solve a self-consistent Schwinger-Dyson equation to estimate the size of an excitonic gap. 
\section*{Results}
The general workflow of our study can be summarized as follows.
\begin{enumerate}
    \item \textbf{Machine learning.} We trained a neural network (the continuous-filter  convolutional  neural  network scheme - SchNet \cite{schutt2017schnet}) on 24,134 band gaps of non-magnetic materials taken from the Organic Materials Database - OMDB \cite{borysov2017organic}. We applied the model to 202,117 materials containing carbon and hydrogen stored in the crystallographic open database - COD \cite{gravzulis2011crystallography}.
    \item \textbf{Band structure calculations.} We select 414 materials where the band gap is predicted small, but nonzero ($0.01~\text{eV}\leq\Delta\leq 0.4$~eV) and perform medium accuracy ab initio calculations using VASP incorporating the effect of spin-orbit interaction (SOI). Note that all calculations stored within the OMDB were peformed without SOI. Out of the 414 materials we found promising features in the band structures for 9 materials. For these 9 materials we performed high-accuracy VASP calculations taking into account structural optimization and SOI. We found the organic charge transfer salts (EDT-TTF-I$_2$)$_2$(DDQ)$\cdot($CH$_3$CN) and TSeF-TCNQ and a bis-1,2,3-dithiazolyl radical conductor which exhibit a semimetallic phase. Based on symmetry and chemistry we determine the relevant mechanisms to protect the nodal features.
    \item \textbf{Effective models and excitonic gap.} We construct an effective model for (EDT-TTF-I$_2$)$_2$(DDQ)$\cdot($CH$_3$CN) and TSeF-TCNQ using qsymm \cite{varjas2018qsymm} and GTPack \cite{gtpack1,gtpack2}. We solve a self-consistent excitonic gap equation, assuming a $s$-wave gap.
\end{enumerate}
We will discuss the outcome of the three steps in more detail in the following.
\subsection{Machine Learning}
\begin{figure}[b!]
    \centering
    \includegraphics[width=0.45\textwidth]{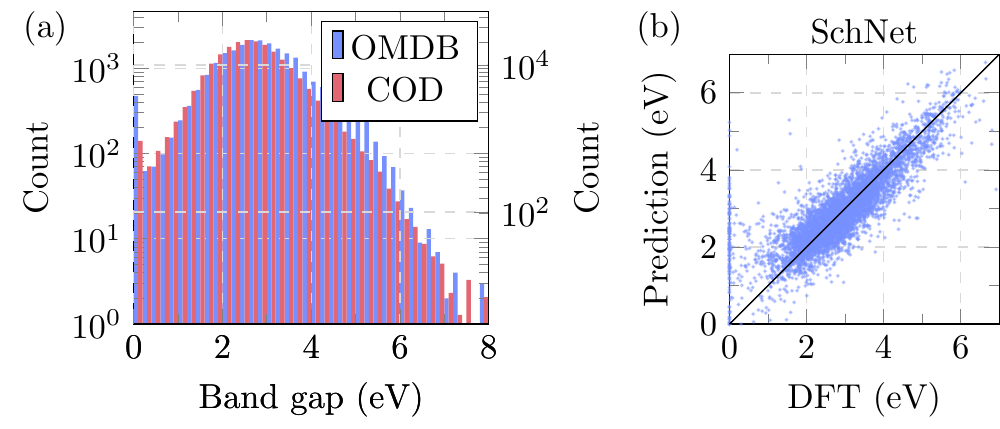}
    \caption{Machine learning predictions of electronic band gaps using the continuous-filter convolutional neural network scheme -- SchNet. (a) band gap statistics of the initial OMDB data ($\approx 2\times10^4$ materials, count on the left) and the predicted gaps ($\approx 2\times10^5$ materials, count on the right); (b) performance of SchNet on the test set.}
    \label{methods:schnet}
\end{figure}

According to the OMDB, organic crystals show a mean ab initio band gap of $\approx 3$ eV with a standard deviation of $1$ eV \cite{borysov2017organic}. Fig. \ref{methods:schnet}(a) shows a comparison of the band gap distribution of the training set with the band gap distribution obtained using machine learning (ML). While the amount of data is much bigger for the materials taken from the COD, the general shape of the histogram of calculated and predicted gaps agrees well, i.e., the ML model successfully reproduced the band gap statistics. Due to the highly complex structures of organic crystals and the relatively small data set, our trained ML model has a large mean absolute error (MAE) of 0.406 eV. Compared to the average band gap of $\approx$ 3 eV, this value represents a sufficient accuracy. However, as we are interested in the tiny gap regime, the order of the MAE is similar to our acceptance range $0.01~\text{eV}\leq\Delta\leq 0.4$ eV. Nevertheless, the general guidance of our model was sufficient to identify a few final example materials. We note that a more sophisticated prediction scheme reaching a high accuracy on small and complex data sets would significantly advance the outcome of our approach in the future. 
\subsection{Band Structure Calculations}
\begin{figure*}
\centering
\includegraphics[width=\textwidth]{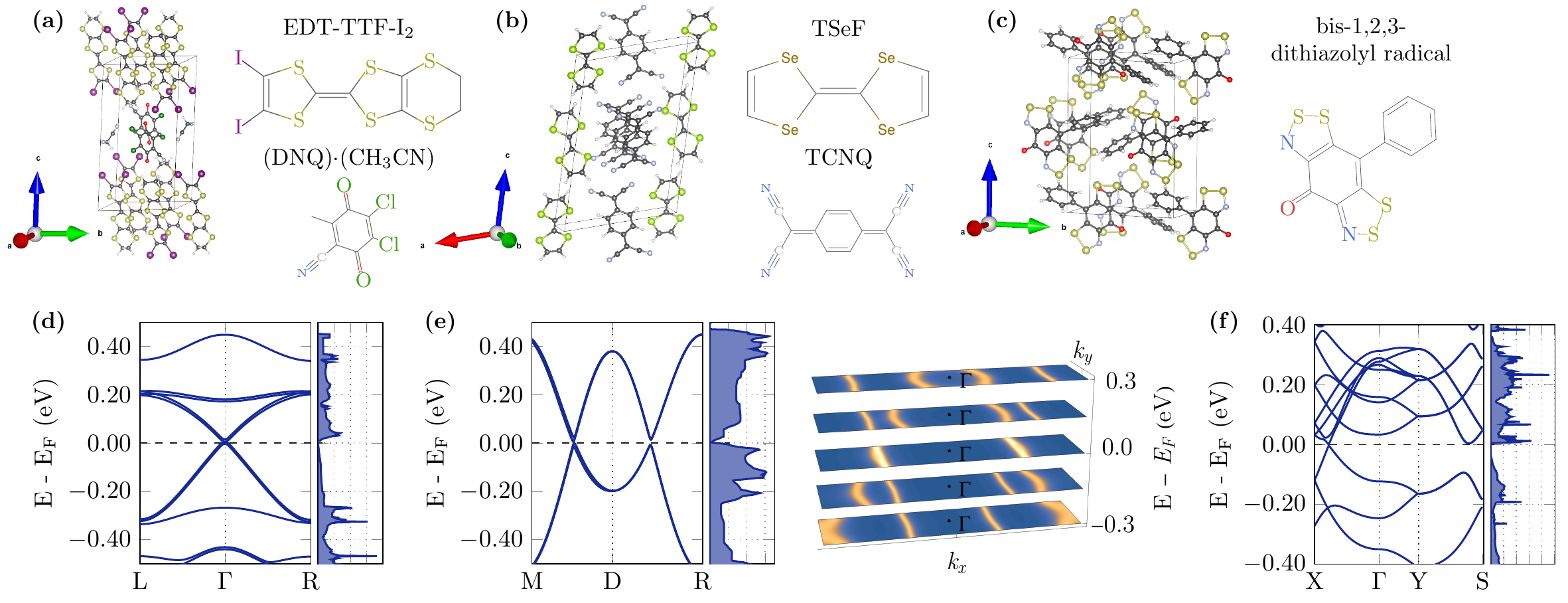}
\caption{(a)-(c) crystal and molecular structures of (EDT-TTF-I$_2$)$_2$(DDQ)$\cdot($CH$_3$CN) (COD-ID: 1508451), TSeF-TCNQ (COD-ID: 2101483) and bis-1,2,3-dithiazolyl (COD-ID: 7106265). The corresponding band structures are shown in (d)-(f). (d) (EDT-TTF-I$_2$)$_2$(DDQ)$\cdot($CH$_3$CN) has a four-fold degenerate Dirac node at the Brillouin zone center ($\Gamma$). (e) (TSeF-TCNQ) shows pairs of line nodes passing through the Brillouin zone. (f) bis-1,2,3-dithiazolyl has a Weyl node along the path $\overline{\Gamma \text{X}}$ and corresponding symmetry partners for the ferromagnetic phase.  \label{main_bands}}
\end{figure*}

We verified the effect of SOI on the 414 predicted tiny gap materials. Our calculations revealed nine materials which exhibit nodes in their electronic band structure close to the Fermi level (respective COD-IDs \cite{gravzulis2011crystallography} are 1508451, 2100065, 2101483, 4114836, 4324376, 4334234, 4506562, 7014584, 7106265; see supplementary material). After performing a structural optimization, only 3 materials remain candidates for organic molecular semimetals: the organic charge transfer salts (EDT-TTF-I$_2$)$_2$(DDQ)$\cdot($CH$_3$CN) (1508451) and TSeF-TCNQ (2101483) and the bis-1,2,3-dithiazolyl radical (7106265) (COD-IDs given in brackets). The crystal and electronic structures are shown in Fig. \ref{main_bands}.

(EDT-TTF-I$_2$)$_2$(DDQ)$\cdot($CH$_3$CN) (bis(3,4-diiodo-3',4'-ethyleneditio-tetrathiafulvalene), 
2,3-dichloro-5,6- dicyanobenzoquinone,acetenitrile) crystallizes in a triclinic lattice with space group P$\overline{1}$ (2). The synthesis and characterization is described in Ref. \cite{lieffrig2012halogen}, where indications towards (EDT-TTF-I$_2$)$_2$(DDQ)$\cdot($CH$_3$CN) being a semiconductor with a conductivity gap of 1220~K ($\approx 105$~meV) were reported. 

Synthesis and characterization of the organic charge transfer salt 2,2',5,5'-tetraselenafulvalene--7,7,8,8-tetracyano-p-quinodimethane (TSeF-TCNQ) can be found in Refs. \cite{corfield1996structure,engler1977organic}. The material crystallizes in the monoclinic space group P$2_1/c$ (14), where TSeF and TCNQ molecules arrange in a segregated stack structure, known to mediate conductivity in organic salts \cite{kistenmacher1974crystal}. The material exhibits a metallic high-temperature phase with a metal to insulator transition at temperatures of $\approx 40$~K \cite{scott}. 
    
The ABABAB $\pi$-stacked bis-1,2,3-dithiazolyl radical was synthesized by Yu \textit{et al.} \cite{yu2011first}. It crystallizes in the non-centrosymmetryc space group P$2_12_12_1$ (19) and orders in a canted antiferromagnetic structure with $T_{\text{N}}\approx 4.5$~K. It was reported to undergo a spin-flop transition to a field-induced ferromagnetic state at 2~K and a magnetic field strength of $H \approx $ 20 $k$Oe.

For all materials the calculated crossing points occur at the Fermi level with simultaneously vanishing DOS. The selected $\vec{k}$-path follows the convention of pymatgen \cite{ong2013python}. (EDT-TTF-I$_2$)$_2$(DDQ)$\cdot($CH$_3$CN) shows a crossing of 4 bands at the Brillouin zone center ($\Gamma$; Fig. \ref{main_bands}(d)) exhibiting a tiny splitting of two two-fold band degeneracies away from $\Gamma$ due to a small magnetization of two central carbon atoms with $\approx$ 0.65 $\mu_B$. For (TSeF-TCNQ) we observe 8-fold degenerate Dirac nodes along the paths $\overline{\mathrm{MD}}$ ($\mathrm{M}=(0.5,0.44,0.6)$, $\mathrm{D}=(0.5,0.0,0.5)$; in units of the reciprocal basis vectors) and $\overline{\mathrm{DR}}$ ($\mathrm{D}=(0.5,0.0,0.5)$, $\mathrm{R}=(0.5,0.5,0.5)$) as well as at the corresponding symmetry partners. These nodes belong to line nodes passing through the Brillouin zone as shown in the density plot of Fig. \ref{main_bands} (e). Our DFT calculations give a zero magnetization for all sites within the (TSeF-TCNQ) unit cell. For the $\pi$-stacked bis-1,2,3-dithiazolyl radical we find a 2-fold degenerate Weyl node within the ferromagnetic phase along the path $\overline{\mathrm{\Gamma X}}$ ($X=(0.5,0.0,0.0)$, Fig. \ref{main_bands} (f)). A comparison with the metallic band structure of the nonmagnetic and the fully-gapped band structure of an antiferromagnetic phase can be found in the supplementary material.

\subsubsection{Protection of the nodes}
\begin{figure}
    \centering
    \includegraphics[width=0.45\textwidth]{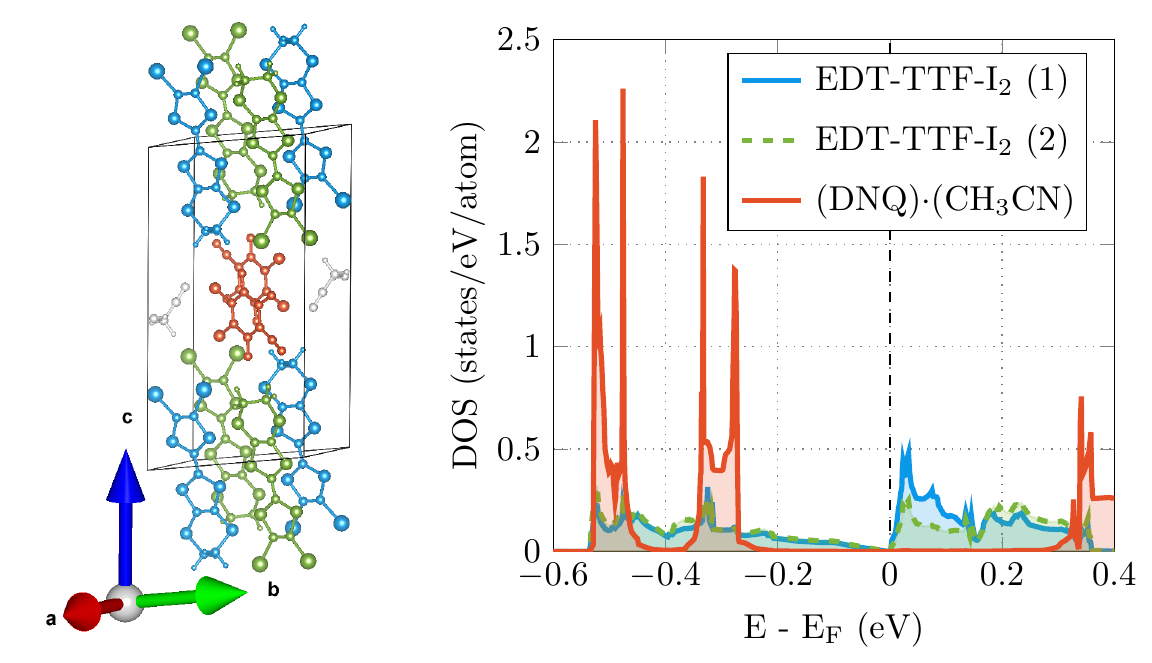}
    \caption{Molecule resolved density of states of (EDT-TTF-I$_2$)$_2$(DDQ)$\cdot($CH$_3$CN). While there are four (two) EDT-TTF-I$_2$ ((DDQ)$\cdot($CH$_3$CN)) molecules in the unit cell, inversion symmetry leads to a pairwise equivalent DOS. The main contribution around the Fermi level is provided by EDT-TTF-I$_2$.}
    \label{EDTpartialDos}
\end{figure}

To understand the nature of the crossings observed in the electronic band structures we are going to discuss the symmetry protection of the nodes. In Fig. \ref{EDTpartialDos} we show the molecule resolved partial DOS of (EDT-TTF-I$_2$)$_2$(DDQ)$\cdot($CH$_3$CN). While there are in total four (EDT-TTF-I$_2$)$_2$ molecules and two DDQ and CH$_3$CN molecules in the unit cell, the inversion symmetry present in the crystal enforces pairwise degenerate contributions to the DOS. As can be verified, the main contribution to the DOS around the Fermi energy stems from (EDT-TTF-I$_2$)$_2$. Due to the specific stacking structure, molecules in charge transfer salts are known to undergo a transition into a dimerized electronic state, where molecular orbitals of pairs of molecules bind significantly stronger to each other than to other molecules in the crystal \cite{lieffrig2012halogen, dimer1, dimer2}. In other words, we can introduce three energy scales: i) the hopping $\tau_{\alpha\beta}$ of electrons between atoms $\alpha,\beta$ within a molecule; ii) the hopping $t_{\mu\nu}$ of electrons between molecules $\mu,\nu$ within a molecular dimer; iii) the hopping $s_{i j}$ of electrons between different dimers $i,j$. In the limit of $s_{ij}\ll t_{\mu\nu} \ll \tau_{\alpha\beta}$, the problem can be separated. Assume that the Hamiltonian describing a EDT-TTF-I$_2$ reveals an electronic ground state with an $s$-like molecular orbital. Then, the Hamiltonian describing the dimerization can be written as $\op{H} = \tau \op{\Psi}
^\dagger \mat{\sigma}_x \op{\Psi}$ with $\op{\Psi} = \left( \op{c}_1,\op{c}_2\right)$, where $\op{c}_i^\dagger$ ($\op{c}_i$) creates (annihilates) an electron in molecule $i$. Hence, the two eigenstates are given by $\phi_\pm=\frac{1}{\sqrt{2}}\left(\op{c}_1 \pm \op{c}_2\right)$ with the respective energies $\pm \tau$. In the crystalline unit cell these dimer states can be used to construct a basis where the $\phi_\pm$ are even (+) and odd (-) with respect to inversion symmetry. Furthermore, taking the center of masses of the (EDT-TTF-I$_2$)$_2$ dimers, the resulting lattice (without (DDQ)$\cdot($CH$_3$CN)) can be approximated by a reduced lattice with one dimer site per unit cell, where the lattice constant in $\vec{a}$ direction is decreased by a factor of $\frac{1}{2}$. This decrease of the unit cell also introduces an effective half-filling as the initial space group symmetry is $P\overline{1}$, which according to its associated Bieberbach manifold is half-filled for an odd number of electrons \cite{watanabe2015filling}. The half-filling together with the effective bipartite lattice introduces (besides parity and time-reversal) a particle-hole symmetry. Considering a four-band model with two orbital and two spin degrees of freedom we obtain the following effective $\vec{k}\cdot\vec{p}$ Hamiltonian around the $\Gamma$ point (see method section for details),
\begin{equation}
    \mat{H}_{\text{eff}} \approx \sum_{i=1}^3 \left[ a_i \tau_1\sigma_1 k_i + b_i \tau_1\sigma_2 k_i\right].
    \label{EDTTTF_effectiveHam}
\end{equation}
The Hamiltonian reveals a symmetry protected four-fold degenerate Dirac node at the center of the Brillouin zone. By lattice symmetry this node does not carry any topological charge according to the Nielsen-Ninomiya theorem \cite{nielsen1981absence,nielsen1981absence2}.

\begin{figure}
    \centering
    \includegraphics[width=0.45\textwidth]{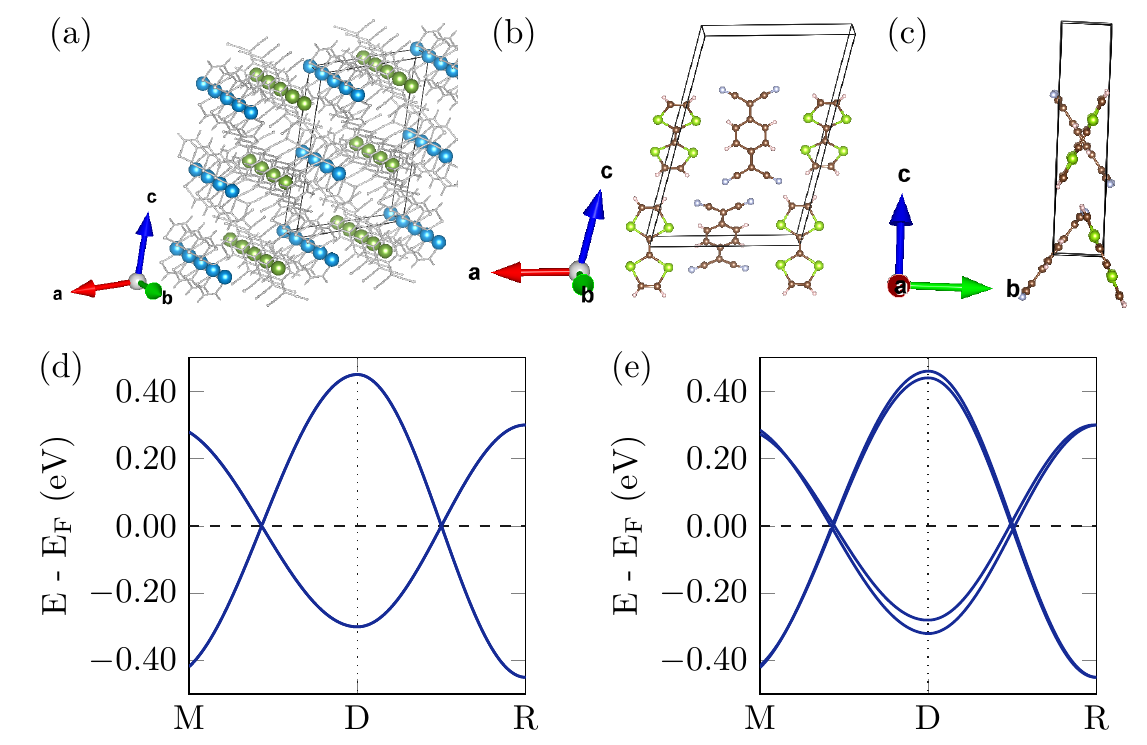}
    \caption{\change{Effective model for weakly interacting chains in TSeF-TCNQ. (a) center of masses of TSeF and TCNQ in the unit cell, forming one-dimensional chains in $\vec{b}$ direction. (b) and (c) illustrates the tilted stacking of molecules leading to a vanishing hopping amplitude between different chains. (d) and (e) show effective band structures without and with weak interchain hopping. }}
    \label{TSFpartial}
\end{figure}

In general, the underlying space group symmetry for (TSeF-TCNQ) (P$2_1/c$) does not allow for the high degeneracy of the nodal line observed. The nature of the crossing stems from the quasi one-dimensional nature of the charge transfer salt due to the specific molecular stacking. In Fig. \ref{TSFpartial} (a) we show the center of mass coordinates of the involved molecules which form weakly interacting one-dimensional chains. Each chain can be approximated to have a dispersion relation of $E_{\mu \sigma} = s_0^\mu + 2 s_1^\mu \cos{\vec{a}_2\cdot\vec{k}}$, with $\mu$ being the molecule index distinguishing between TSeF and TCNQ and $\sigma$ denoting the spin (note that the dispersion relation itself is independent of the spin). As there are two TSeF chains as well as two TCNQ chains present in the crystal, we observe two fourfold degenerate bands (2 chains $\times$ 2 spins per molecule) which are allowed to cross in a plane if no hybridization is taken into account (see Fig. \ref{TSFpartial} (d)). From the tilted stacking of molecules in the different chains (shown in Fig. \ref{TSFpartial} (b)+(c)) it becomes apparent that the involved hopping of electrons between different chains (interchain coupling) has to be weaker by several order of magnitude than the intrachain coupling (see also fitted parameters in methods section). In the basis $\op{\Psi} = \left(\op{c}_{\text{TSeF-1} \sigma}, \op{c}_{\text{TCNQ-1} \sigma}, \op{c}_{\text{TSeF-2}\sigma}, \op{c}_{\text{TCNQ-2} \sigma} \right)^T$ and allowing for a hopping between chains we formulate an effective Hamiltonian of the form
\begin{equation}
\mat{H}_\sigma(\vec{k}) =
    \left(
    \begin{array}{cccc}
        E_{\text{TSeF} \sigma} & \Delta_1 & \Delta_2 & 0 \\
        \Delta_1 & E_{\text{TCNQ} \sigma} & 0 & \Delta_3 \\
        \Delta_2 & 0 & E_{\text{TSeF} \sigma} & \Delta_1  \\
        0 & \Delta_3 & \Delta_1 & E_{\text{TCNQ} \sigma} 
    \end{array}
    \right),
\end{equation}
where we use $\Delta_1 = t_1 \cos\left(\frac{1}{2}\vec{a}_1\cdot\vec{k}\right)$, $\Delta_{2,3} = t_{2,3} \cos\left(\frac{1}{2}\left(\vec{a}_2+\vec{a}_3\right)\cdot\vec{k}\right)$. An example band structure for $t_1\neq t_2 \neq t_3 \approx 10^{-2} s_{1}^{\mu}$ is shown in Fig. \ref{TSFpartial} (e) which reproduces the ab initio band structure of Fig. \ref{main_bands} (e) effectively. A slightly more advanced effective Hamiltonian is described in the methods section.

\begin{figure}
    \centering
    \includegraphics[width=0.45\textwidth]{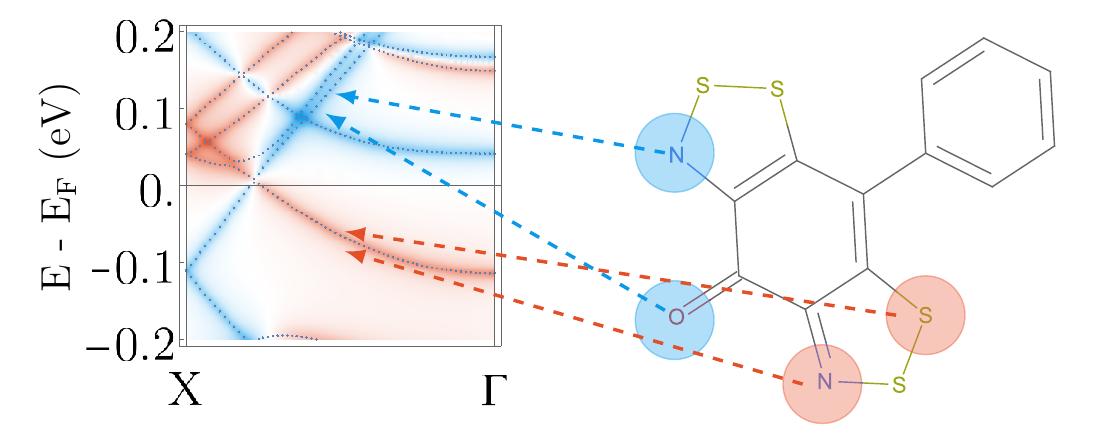}
    \caption{Atomic species contributing to the formation of Weyl nodes in bis-1,2,3-dithiazolyl.}
    \label{dithiaWeyl}
\end{figure}

The radical bis-1,2,3-dithiazolyl exhibits Weyl nodes for the low-temperature high-field ferromagnetic phase. Hence, time-reversal symmetry is broken and the crystal reflects a magnetic space group symmetry, depending on the magnetization direction. Assume we choose the magnetization in $x$-direction then the corresponding group is given by $P2_12_1'2_1'$ (\# 19.27). To verify that the observed nodes are Weyl nodes we calculated the atom and orbital resolved weights to the band structure. We observe that the two bands forming the node belong to different atoms in the molecules within the unit cell. The main contributions stem from either the unsaturated nitrogen and the oxygen atom or the saturated nitrogen and one sulfur atom as shown in Fig. \ref{dithiaWeyl}. Hence, we conclude that the bands belong to different orbital subspaces allowing for a crossing. In symmetry terms the two orbitals correspond to bands with different eigenvalues to the only unitary symmetry element present, $(C_{2x}, 1/2,1/2,0)$, along the invariant line $k_x$. Hence, a 2-band $\vec{k}\cdot\vec{p}$ Hamiltonian at a crossing point has to be invariant under the Pauli matrix $\mat{\tau}_x$, $\mat{H}(k_x,k_y,k_z) = \mat{\tau}_x \mat{H}(k_x,-k_y,-k_z) \mat{\tau}_x$. This results in an allowed low energy Hamiltonian $\mat{H}(\vec{K}
^\pm+\vec{k}) \approx \pm t_x k_x \mat{\tau}_x + t_y k_y \mat{\tau}_y + t_z k_z \mat{\tau}_z$, at two points $\vec{K}^\pm$ along the $x$-axis. Both points reveal a monopole charge of $\nu^\pm = \operatorname{sign}\left[\pm t_x t_y t_z\right]$.

\subsection{Effective Models and Excitonic Gap}
To explain the discrepancy between DFT semimetallic phases and experimentally observed semiconductivity in (EDT-TTF-I$_2$)$_2$(DDQ)$\cdot($CH$_3$CN) and (TSeF-TCNQ) we argue that both systems are likely to undergo an excitonic instability. Compared to known inorganic Dirac or Weyl semimetals, the band width of all three materials discussed here is smaller by at least one order of magnitude. As the decreased band width induces a small Dirac velocity (small kinetic energy) of the elementary excitations of the system, we verify that a quasiparticle interaction term becomes dominant. We estimate the size of the effect using effective band structure models. For (EDT-TTF-I$_2$)$_2$(DDQ)$\cdot($CH$_3$CN) we extended the model given in \eqref{EDTTTF_effectiveHam} to a lattice periodic version ($x \rightarrow \sin(x)$). For (TSeF-TCNQ) we construct an eight band model incorporating two spin and four orbital degrees of freedom (two belonging to molecular orbitals from TSeF and two belonging to TCNQ). The allowed dispersion relation of the model is restricted by the symmetry generators: parity, two-fold rotation about $x$-axis, time-reversal symmetry (details given in the methods section).

To approximate the size of the instability, we solve a BCS-like excitonic gap equation, similar to Refs \cite{khveshcenko, triola2017excitonic, pertsova2018excitonic}. Note that given the numerous degrees of freedom, multiple excitonic gap symmetries might occur, such as inter- or intranode instabilities with or without breaking of time-reversal or other lattice symmetries \cite{pertsova2018excitonic, Sukhachov2019}. We focus on an $s$-wave gap. Details on the calculation are given in the methods section.

\begin{figure}[t!]
    \centering
    \includegraphics[width=0.49\textwidth]{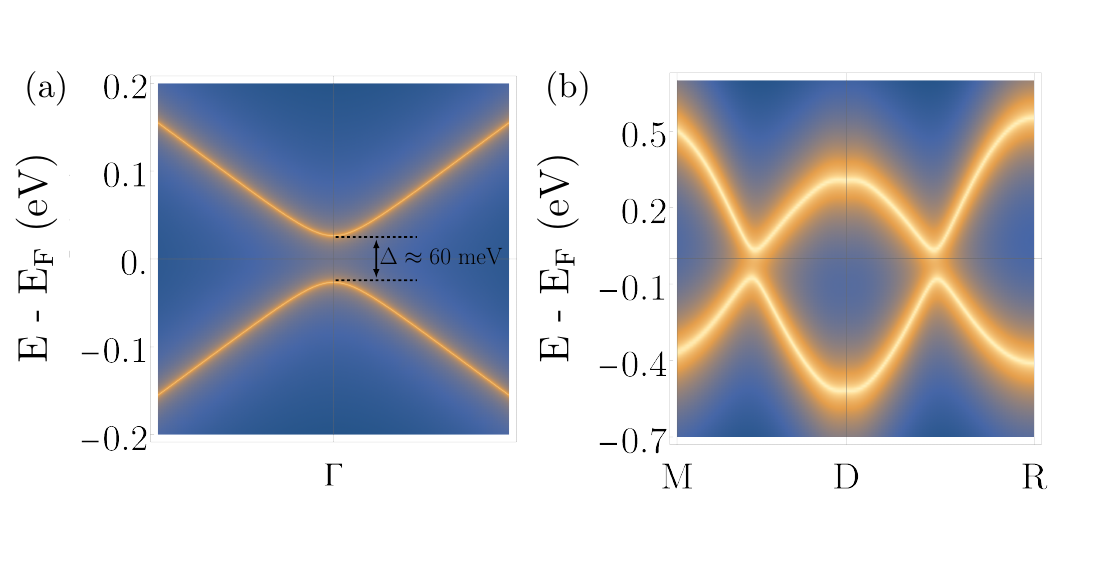}\vspace{-0.5cm}
    \caption{\change{Calculated approximations for an excitonic gap at zero temperature for (a) (EDT-TTF-I$_2$)$_2$(DDQ)$\cdot($CH$_3$CN) and (b) (TSeF-TCNQ), obtained by numerically solving a BCS-like gap equation within the full Brillouin zone.}}
    \label{excitonic_inst}
\end{figure}

We obtained an excitonic gap of $\approx 60$ meV for (EDT-TTF-I$_2$)$_2$(DDQ)$\cdot($CH$_3$CN) at $T\rightarrow 0$ K. This value agrees with the experimentally determined gap of $\approx 105$ meV. We argue that despite (EDT-TTF-I$_2$)$_2$(DDQ)$\cdot($CH$_3$CN) being a regular semiconductor, it is an excitonic insulator with an electronic structure shown in Fig. \ref{excitonic_inst}(a). We repeat the approach for (TSeF-TCNQ) and obtain a slightly higher value of $\approx 150$ meV (Fig. \ref{excitonic_inst}(b)). In contrast to the Dirac semimetal (EDT-TTF-I$_2$)$_2$(DDQ)$\cdot($CH$_3$CN), (TSeF-TCNQ) has two degenerate line nodes which gap out while the excitonic gap is formed. To verify the experimentally observed metal-insulator transition at 40~K would require a full temperature-dependent screening which is beyond our present study.
\section{Methods}
\subsection{Machine Learning}
The ML model is trained on a dataset of 24,134 ab initio band gaps of non-magnetic organic crystals stored in the organic materials database - OMDB \cite{borysov2017organic,olsthoorn2019band}. The dataset is divided into a training, validation and test set of 15000, 3000 and 6134 materials, respectively. We used the continuous-filter convolutional neural network scheme - SchNet \cite{schutt2017schnet} with a batch size of 32, a cutoff radius of 5.0 \AA, 32 features, 3 interaction blocks, a learning rate of $10^{-4}$, and 50 Gaussians to expand atomic distances (see Ref. \cite{olsthoorn2019band} for parameter choice).
The initial band gap data exhibits a Wigner-Dyson like shape as shown in Fig. \ref{methods:schnet}(a), having a mean of $\approx 2.9$ eV and a standard deviation of $\approx 1.1$ eV. Our trained ML model shows a mean absolute error (MAE) of 0.406 eV and a RMSE of 0.602 eV which is interpreted as an accuracy of $\approx 90\%$. Note that the underlying data is rather small ($\approx 2\times 10^{4}$ materials) and highly complex (average of 85 atoms per unit cell). The trained ML model is able to reproduce the band gap statistics on a set of 202,117 organic crystals ($\approx 2\times 10^{4}$ materials) taken from the COD \cite{gravzulis2011crystallography}, as shown in Fig. \ref{methods:schnet}(a). These predictions also fit a Wigner-Dyson distribution $\sim x^{5.62} e^{-0.45 x^2}$. The performance of our model on the test set is shown in Fig. \ref{methods:schnet}(b).

\subsection{ab initio Calculations}
We performed ab initio calculations in the framework of the density functional theory (DFT) using a projector augmented-wave method as implemented in the Vienna ab initio simmulation package VASP \cite{vasp}. Structures are taken from the  COD \cite{gravzulis2011crystallography} and transformed into VASP input using pymatgen \cite{ong2013python}. During the self-consistent calculation of the electron density and the DOS we used a $\Gamma$-centered mesh with a $\vec{k}$-mesh density of 800 points per $\AA^{-3}$ for the quick materials scan of 414 candidate materials and a $\vec{k}$-mesh density of 1500 points per $\AA^{-3}$ for the refined calculations of 9 prospective semimetals. For the latter we performed an optimization of the atomic positions using a conjugate gradient algorithm. All calculations were performed with SOI and the exchange correlation functional was approximated by the strongly constrained and appropriately normed semilocal density functional SCAN \cite{scan2} including van-der-Waals-corrections using the Tkatchenko-Scheffler method with iterative Hirshfeld partitioning \cite{tkatchenko2009accurate,bucko2013improved}. The cut-off energy of the plane wave expansion was chosen to by 600 eV. \nocite{momma2011vesta}
\subsection{Effective Hamiltonians}
We generated effective Hamiltonians for (EDT-TTF-I$_2$)$_2$(DDQ)$\cdot($CH$_3$CN) and TSeF-TCNQ. To describe the four-fold degeneracy at the $\Gamma$ point observed for (EDT-TTF-I$_2$)$_2$(DDQ)$\cdot($CH$_3$CN), we construct a model with two orbital ($\tau_i$) and two spin ($\sigma_i$) degrees of freedom,
\begin{equation}
    \mat{H}_{\text{eff}} = \sum_{i=0}^3\sum_{j=0}^3 f_{ij}(k_x,k_y,k_z) \tau_i\sigma_j.
\end{equation}
Here, each molecule in the unit cell contributes one molecular orbital. Neglecting higher and lower energy bands, a four-fold degeneracy is obtained assuming an even- and an odd-parity molecular orbital, and add respective lattice symmetries (space group $\mathrm{P}\overline{1}$): parity $\op{P} = \tau_3\sigma_0$ ($\vec{k}\rightarrow -\vec{k}$), time-reversal $-\mathrm{i} \sigma_2 \op{K}$ ($\vec{k}\rightarrow -\vec{k}$; $\op{K}$ being the complex conjugation), and an emergent particle-hole symmetry $\op{C}= \sigma_2$. Up to linear order we obtain
\begin{equation}
    \mat{H}_{\text{eff}} \approx \sum_{i=1}^3 \left[ a_i \tau_1\sigma_1 k_i + b_i \tau_1\sigma_2 k_i\right].
\end{equation}
The best fit to the ab initio calculated band structure resulted in the parameters $a_1 = 0.47$ eV, $a_2 = -0.33$ eV, $a_3 = 0.11$ eV, $b_1 = 0.16$ eV, $b_2 = -0.38$ eV, $b_3 = 0.15$ eV.

The nodes for TSeF-TCNQ occur in the interior of the Brillouin zone. Therefore, we construct a lattice-periodic model with 8 bands (4 orbital-, 2 spin degrees of freedom). The orbital degrees of freedom are obtained from respective permutations of the two TSeF and two TCNQ molecules within the primitive unit cell. We represent the generators of the factor group $\mathcal{G}/\mathcal{T} \simeq C_{2h}$ ($\mathcal{G}$ is the space group P2$_1$/c, $\mathcal{T}$ is the corresponding group of pure lattice translations) as follows, $\op{P} = \mat{P}(1,2,3,4)\times \sigma_0$ ($\vec{k}\rightarrow -\vec{k}$), $\op{C}_{2x} = \mat{P}(2,1,4,3)\times \mathrm{i} \sigma_1$ ($k_x\rightarrow k_x$, $k_y\rightarrow -k_y$, $k_z\rightarrow -k_z$) and add time-reversal symmetry as before. Here $\mat{P}(\text{permutation})$ denotes a $4\times4$-dimensional permutation matrix. Note that each symmetry element comes with a corresponding double group partner. The total Hamiltonian is written as $\mat{H}=\mat{H}^{\text{orbital}}\times\mat{H}^{\text{spin}}$.
We generate $\mat{H}_{\text{orbital}}$ in the basis ($\left|\Psi_{\text{TSeF1}} \right>$, $\left|\Psi_{\text{TSeF2}} \right>$, $\left|\Psi_{\text{TCNQ1}} \right>$, $\left|\Psi_{\text{TCNQ2}} \right>$) using 
\begin{align}
&H^{\text{orbital}}_{mn} = t_{mn} \exp\left(\mathrm{i} \vec{k} \cdot \vec{\delta}_{mn}\right) \notag\\
&H^{\text{orbital}}_{mm} = \sum s^m_i \exp\left(\mathrm{i} \vec{k} \cdot \vec{a}_i\right),
\end{align}
with $m,n=\text{TSeF1},\text{TSeF2},\text{TCNQ1},\text{TCNQ2}$,
 $\vec{\delta}_{mn}$ being the vector connecting molecules $m$ and $n$ and $\vec{a}_i$ being the real space basis vectors. The full Hamiltonian is obtained by applying all symmetry operations and dividing by the total number of symmetry elements, i.e., twice the order of the factor group when all combinations incorporating time-reversal symmetry are taken into account. We implemented a Monte-Carlo scheme to fit the model to our ab initio band structure with the best fit being $t_{\text{TSeF1,TSeF2}} = 0.008$ eV, $t_{\text{TCNQ1,TCNQ2}}=-0.025$ eV, $ t_{\text{TSeF1,TCNQ1}}=0.023$ eV, $t_{\text{TSeF1,TCNQ2}}=0.047$ eV, $s^{\text{TSeF1}}_1 = -0.033$ eV, $s^{\text{TSeF1}}_2=-0.023$ eV, $s^{\text{TSeF1}}_3=1.057$ eV, $s^{\text{TCNQ1}}_1=-0.704$ eV, $s^{\text{TCNQ1}}_2 = 0.080$ eV, $s^{\text{TCNQ1}}_3=-0.093$ eV.

\subsection{Excitonic Instability}
We model quasiparticle-quasihole excitations of the material in terms of the following interaction process, where incoming (outgoing) solid lines belong to fermionic annihilation (creation) operators $\op{\Psi}_{\vec{q}}$ ($\op{\Psi}^\dagger_{\vec{q}}$) and the dashed line to a screened scalar interaction $V(\vec{q})$.  
\begin{center}
\includegraphics[height=2cm]{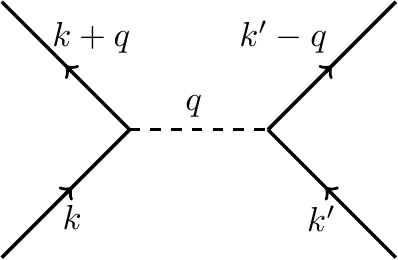}
\end{center}
The Hamiltonian is given by
\begin{equation} 
    \op{H} = \op{H}_0 
    + \frac{1}{2} \sum_{a\nu} \sum_{\vec{k} \vec{k}' \vec{q}} V(q) \op{\Psi}^\dagger_{\nu,\vec{k}+\vec{q} a} \op{\Psi}^\dagger_{\nu,\vec{k}'-\vec{q} \overline{a}} \op{\Psi}_{\nu,\vec{k}' \overline{a}} \op{\Psi}_{\nu,\vec{k}a}.
\end{equation}
Here, $\nu$ is a corresponding quantum number and $a=\pm$ ($\overline{a}=-a$) denotes electron and hole states. $\op{H}_0$ denotes the non-interacting effective Hamiltonian of the system. 
We derive an excitonic gap equation using a Green function approach similar to Refs \cite{khveshcenko, triola2017excitonic, pertsova2018excitonic} and define $G_{\nu,\pm}(\vec{p},\tau-\tau') = -\left<\op{T} \op{\Psi}_{\nu, \vec{p}\pm}(\tau) \op{\Psi}^\dagger_{\nu, \vec{p}\pm}(\tau') \right>
$ and $F_{\nu,\pm}(\vec{p},\tau-\tau') = -\left<\op{T} \op{\Psi}_{\nu, \vec{p}\pm}(\tau) \op{\Psi}^\dagger_{\nu, \vec{p}\mp}(\tau') \right>$.
We proceed by calculating the equations of motion for $G$ and $F$ using the Heisenberg formalism and decomposing the four-point functions in terms of $G$ and $F$ \cite{mahan2013many}. We impose particle-hole symmetry of the electronic band structure $\xi^+_\nu(\vec{k}) = \xi^-_\nu(\vec{k})$ as well as a real excitonic gap function $\Delta_\nu$ and derive the BCS-like gap equation
\begin{equation}
    \Delta_{\nu}(\vec{p}) = -\sum_{\vec{q}} V(\vec{q}-\vec{p}) \frac{\Delta_{\nu}(\vec{q})}{2 E_{\nu,\vec{q}}} \tanh{\left(\frac{\beta E_{\nu,\vec{q}}}{2}\right)},
    \label{BCSeq}
\end{equation} 
which is in agreement with Ref. \cite{khveshcenko}. Here, $E_{\nu,\vec{p}}=\sqrt{\xi_\nu^2(\vec{p}) + \Delta_\nu^2(\vec{p})}$. We construct the exact Brillouin zone using the algorithm of Finney \cite{finney1979procedure} and solve \eqref{BCSeq} using a Liouville-Neumann series, where the integration is performed using a quasi-Monte-Carlo integration scheme \cite{Mathematica}. We calculate the zero temperature excitonic gap $\tanh{\left(\frac{\beta E_{\nu,\vec{q}}}{2}\right)} \rightarrow 1$ and approximate the interaction by a screened Coulomb interaction $V(\vec{q}) = - \frac{4\pi}{\vec{q}^2 + \vec{k}^2},
$
with $\vec{k}$ being a screening vector generally temperature temperature dependent \cite{khveshcenko}. The screening vector is approximated by the Thomas-Fermi screening, which for Dirac semimetals is given by $\vec{k} = \sqrt{\frac{2 g \alpha}{\pi}} \vec{k}_{\text{F}}$ \cite{pertsova2018excitonic},
with $\vec{k}_{\text{F}}$ being the Fermi-vector and $g$ being the Dirac cone degeneracy. Organic crystals typically exhibit purification during the crystallization process expelling dopants. We assume a clean sample with Fermi-level deviation $\delta \mu = 0.01$ eV.

\section{Conclusion}
Typically, nodal states within the electronic structure of organic crystals are inferred by indirect measurements, e.g., via the resistivity \cite{Liu2016}, optical conductivity \cite{Beyer2016}, and local spin susceptibility \cite{hirata2016observation} as has been done extensively for the pressure induced semimetal phase of (BEDT-TTF)$_2$I$_3$. A direct observation of the electronic structure of organic materials with angle-resolved photoemission spectroscopy (ARPES) is difficult due to: usually tiny crystal sizes limiting signal; insulating behavior leading to charging of the samples; limits in available orientations and problems in preparing well-defined surface terminations. However, ARPES has been performed for organics in selected cases \cite{koller2007intra}. Here, the reported experimental maximum/median/minimum crystal sizes (mm) of the materials described by us are 0.24/0.05/0.04 for (EDT-TTF-I$_2$)$_2$(DDQ)$\cdot($CH$_3$CN), 0.23/0.05/0.02/ for (TSeF-TCNQ), and 0.34/0.04/0.02 for bis-1,2,3-dithiazolyl. While these crystal sizes are tiny, recent progress in the field has decreased the focus area of state-of-the-art ARPES devices tremendously \cite{arpes}, making a measurement of the photo electron spectrum of the three materials possible in the near future. The materials reported here provide a natural platform to investigate the rare excitonic insulator phase as a consequence of strong interaction effects within symmetry protected nodal semimetals.

While several inorganic nodal semimetals are known, the space of organics remains fairly unexplored in this regard. The realm of 3D organic  semimetals is mainly composed of the single example $\alpha$-(BEDT-TTF)$_2$I$_3$ and modifications which exhibit tilted Dirac nodes under pressure ($\approx$ 2.3 GPa) \cite{katayama2006pressure,hirata2016observation} or chemical strain \cite{geilhufe2018chemical}. Hence, the outcome of our ML and ab initio calculations are promising with respect to the identification of novel examples. We note that the choice of the $s$-wave excitonic instability due to strong interactions is only the simplest scenario and was chosen to estimate the size of the effect. However, other forms of gap openings and richer phase diagrams are imaginable. 
\section{Acknowledgements}
We thank Alexander Balatsky, Stanislav Borysov, Pavlo Sukhachov, Vladimir Juri\v ci\' c, and Gayanath Fernando for helpful discussions. We acknowledge funding from the VILLUM FONDEN via the Centre of Excellence for Dirac Materials (Grant No. 11744) and the European Research Council ERC HERO grant. The authors acknowledge computational resources from the Swedish National Infrastructure for Computing (SNIC) at the National Supercomputer Centre at Link\"oping University, the Centre for High Performance Computing (PDC), the High Performance Computing Centre North (HPC2N), and the Uppsala Multidisciplinary Centre for Advanced Computational Science (UPPMAX).

\appendix
\section{Appendix A: Materials Candidates}
\begin{figure*}
    \centering
    \includegraphics{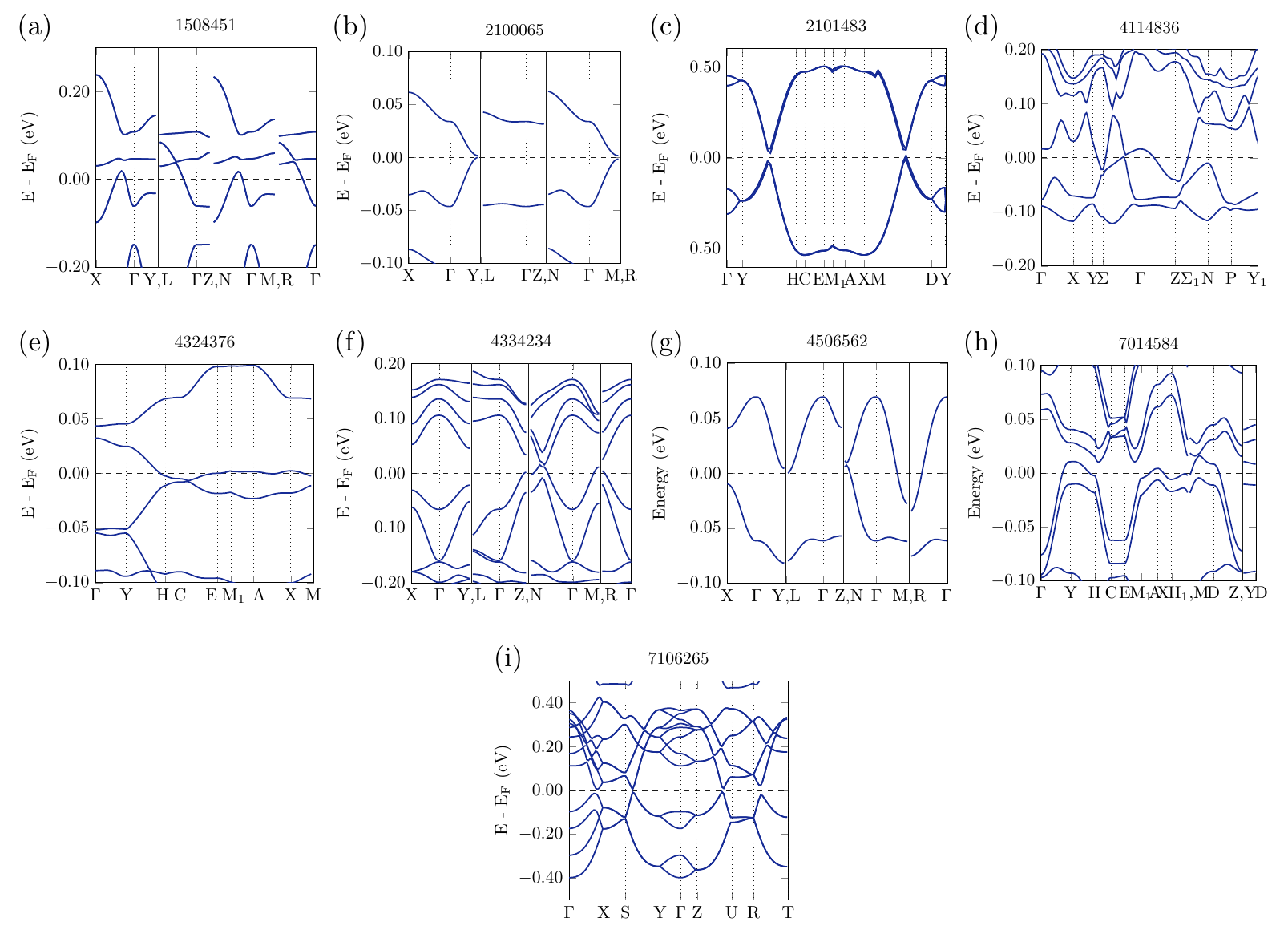}
    \caption{\textbf{Promising materials candidates.} Out of 414 band structure calculations of organic molecular crystals and metal organic frameworks, 9 materials exhibited nodes close to the Fermi level and were selected for revision using a more sophisticated computational approach. The number indicates the COD ID of the material.}
    \label{candidates}
\end{figure*}

Out of the 202,117 materials in the crystallographic open database COD \cite{gravzulis2011crystallography}, a subset of 414 organic materials were predicted to have a small band gap ($0.01 \leq \Delta \leq 0.4$ eV) by our trained machine learning model. We performed coarse ab initio calculations to scan for potential organic semimetals (details on the calculations can be found in the methods section) and find 9 materials candidates which exhibit nodes close to the Fermi level ($E = 0$ eV). The band structures of the nine materials are shown in Fig. \ref{candidates}. Materials are denoted with their COD-IDs. The labeling of the high symmetry points was generated automatically using pymatgen \cite{ong2013python}. Respective space group symmetries are P$\overline{1}$: 1508451, 2100065, 4334234, 4506562; P$2_1/c$: 2101483, 4324376, 7014584; I$\overline{4}2m$: 4114836; P$2_12_12_1$: 7106265. Of the nine materials only the three materials with COD-IDs 1508451, 2101483, and 7106265 still exhibit their nodes after a full structural optimization was performed. 

\section{Appendix B: Magnetic phases of bis-1,2,3-dithiazolyl}
\begin{figure}[t!]
\includegraphics[width=0.49\textwidth]{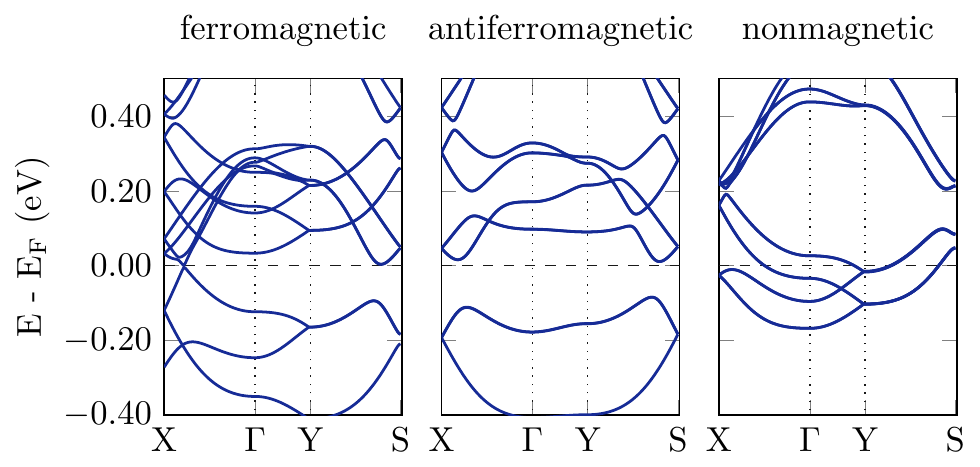}
\caption{Comparison of electronic structure for various magnetic phases of the bis-1,2,3-dithiazolyl radical. }\label{magphases7x}
\end{figure}

The ABABAB $\pi$-stacked bis-1,2,3-dithiazolyl radical was reported to undergo a transition into a canted anti-ferromagnet below a critical temperature of 4.5~K and to undergo a spin-flop transition to a field-induced ferromagnetic state at a temperature of 2~K and a magnetic field strength of $H \approx $ 20 $k$Oe  \cite{yu2011first}. Our calculations revealed Weyl nodes along the path $\overline{\Gamma X}$ in the Brillouin zone for the ferromagnetic phase. To better understand the emergence of the phase and its correspondence to the underlying magnetic ordering we performed similar band structure calculations incorporating full structural optimization for an antiferromagnetic and a non-magnetic ordering (details on the ab initio calculation can be found in the methods section). Even though the nonmagnetic phase exhibits topological nodes along the path $\overline{\Gamma Y}$ the overall behavior is metallic as there is no vanishing density of states at the Fermi level. The topological protection of the nodes for the nonmagnetic phase is a consequence of the underlying orthorhombic symmetry with space group P$2_12_12_1$ where the mechanism is described in Ref. \cite{geilhufe2017data}. The antiferromagnetic phase exhibits a clear band gap of the size of $\approx 0.15$ eV. The comparison of the electronic structure for all three magnetic phases is shown in Fig. \ref{magphases7x}.
\section{Appendix C: Nodes in (EDT-TTF-I)2(TCNQ), COD-ID 4506562}
\begin{figure}[h!]
\includegraphics[]{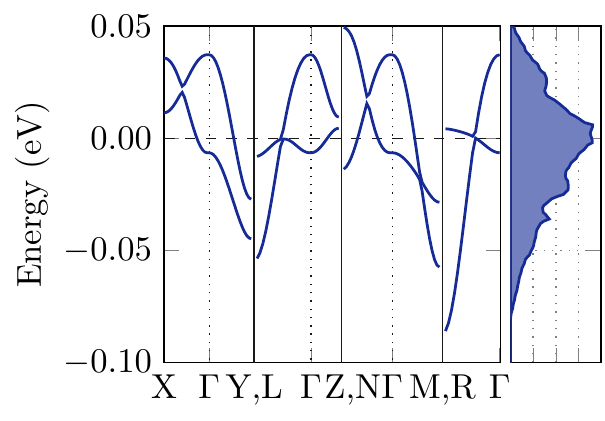}
\caption{Topological nodes in the fully optimized band structure of (EDT-TTF-I)2(TCNQ), COD-ID 4506562.}
\label{bands4506562}
\end{figure}

The material (EDT-TTF-I)2(TCNQ) with COD-ID 4506562 exhibits topological nodes within the Brillouin zone close to the Fermi level as shown in Fig. \ref{bands4506562}. However, the material is not a semimetal as it exhibits a large density of states at the Fermi level.

\section{Appendix D: List of predictions}

We trained a machine learning model based on the continuous-filter convolutional neural network scheme SchNet \cite{schutt2017schnet} on 24,134 ab initio calculated band gaps stored within the organic materials database - OMDB \cite{borysov2017organic} (details can be found in the method section). We applied the successfully trained model on 202,117 crystal structures stored within the crystallographic open database - COD \cite{gravzulis2011crystallography}. All crystal structures considered belong to organic molecular crystals or metal organic frameworks which were synthesized before. Organic materials tend to be large band gap insulators and only 414 materials were predicted to have a band gap of $0.01 \leq \Delta \leq 0.4$ eV, where we explicitly tried to exclude organic metals. The list of predicted band gaps for the 414 materials denoted with their COD-ID is given in Table \ref{predictions}.

\bibliography{references}

\begin{widetext}
\begin{table}
    \centering
    \caption{COD-IDs, band gap $\Delta$, and number of atoms in the primitive unit cell $N_{\text{at}}$ for the subset of 414 COD materials where the band gap was predicted to be small ($0.01 \leq \Delta \leq 0.4$ eV). }
    \label{predictions}
    \tiny
    \begin{tabular}{ccc|ccc|ccc|ccc|ccc|ccc}
        COD ID & $\Delta$ (eV) & $N_{\text{at}}$ & COD ID & $\Delta$ (eV) & $N_{\text{at}}$ & COD ID & $\Delta$ (eV) & $N_{\text{at}}$ & COD ID & $\Delta$ (eV) & $N_{\text{at}}$ & COD ID & $\Delta$ (eV) & $N_{\text{at}}$ & COD ID & $\Delta$ (eV) & $N_{\text{at}}$ \\
        \hline
1100155  &  0.33  &  184  &  2214357  &  0.09  &  172  &  4112733  &  0.24  &  44  &  4327718  &  0.27  &  290  &  7016726  &  0.07  &  192  &  7203122  &  0.03  &  82  \\
1100654  &  0.15  &  204  &  2215187  &  0.27  &  31  &  4112835  &  0.03  &  312  &  4328589  &  0.10  &  348  &  7017188  &  0.06  &  94  &  7203123  &  0.08  &  82  \\
1502567  &  0.15  &  380  &  2215286  &  0.31  &  264  &  4113773  &  0.02  &  312  &  4328637  &  0.26  &  58  &  7018262  &  0.16  &  183  &  7203124  &  0.34  &  82  \\
1502778  &  0.11  &  312  &  2219623  &  0.21  &  232  &  4114183  &  0.05  &  76  &  4329184  &  0.26  &  164  &  7019345  &  0.31  &  248  &  7203125  &  0.31  &  82  \\
1505707  &  0.26  &  176  &  2220473  &  0.30  &  88  &  4114836  &  0.16  &  248  &  4329392  &  0.32  &  444  &  7020987  &  0.39  &  158  &  7203819  &  0.15  &  114  \\
1505708  &  0.27  &  176  &  2222403  &  0.37  &  200  &  4115030  &  0.12  &  68  &  4329599  &  0.36  &  181  &  7024375  &  0.39  &  124  &  7203820  &  0.05  &  114  \\
1506394  &  0.38  &  168  &  2224043  &  0.17  &  56  &  4115033  &  0.32  &  184  &  4329817  &  0.02  &  190  &  7024640  &  0.24  &  392  &  7203821  &  0.13  &  114  \\
1508450  &  0.14  &  68  &  2225643  &  0.05  &  156  &  4115740  &  0.34  &  180  &  4329818  &  0.07  &  288  &  7026747  &  0.10  &  378  &  7204018  &  0.17  &  104  \\
1508451  &  0.32  &  120  &  2231478  &  0.28  &  136  &  4116342  &  0.11  &  156  &  4331203  &  0.39  &  216  &  7027004  &  0.09  &  208  &  7204068  &  0.11  &  43  \\
1508462  &  0.05  &  204  &  2232523  &  0.26  &  48  &  4116343  &  0.21  &  156  &  4331204  &  0.31  &  216  &  7027005  &  0.28  &  176  &  7204069  &  0.06  &  43  \\
1508463  &  0.07  &  204  &  2232886  &  0.02  &  204  &  4116344  &  0.27  &  156  &  4331205  &  0.23  &  216  &  7028070  &  0.27  &  56  &  7204124  &  0.21  &  100  \\
1508464  &  0.33  &  204  &  2233079  &  0.05  &  118  &  4116348  &  0.10  &  156  &  4331206  &  0.24  &  216  &  7028071  &  0.02  &  56  &  7204137  &  0.39  &  106  \\
1508468  &  0.35  &  228  &  2233541  &  0.31  &  216  &  4116732  &  0.39  &  64  &  4331621  &  0.23  &  235  &  7028189  &  0.32  &  106  &  7204163  &  0.33  &  34  \\
1508469  &  0.13  &  228  &  2233927  &  0.25  &  188  &  4117045  &  0.16  &  136  &  4332184  &  0.36  &  150  &  7028313  &  0.35  &  308  &  7204326  &  0.08  &  46  \\
1508470  &  0.17  &  228  &  2235994  &  0.37  &  152  &  4117071  &  0.19  &  156  &  4332331  &  0.09  &  248  &  7031696  &  0.14  &  144  &  7204639  &  0.26  &  204  \\
1512824  &  0.04  &  202  &  2236574  &  0.14  &  108  &  4117380  &  0.32  &  80  &  4333342  &  0.11  &  152  &  7032641  &  0.20  &  164  &  7204757  &  0.10  &  90  \\
1514418  &  0.28  &  304  &  2310537  &  0.20  &  24  &  4118108  &  0.29  &  200  &  4333919  &  0.13  &  106  &  7033203  &  0.28  &  76  &  7205290  &  0.25  &  172  \\
1516182  &  0.27  &  104  &  4000046  &  0.38  &  98  &  4118948  &  0.22  &  72  &  4333923  &  0.32  &  424  &  7035665  &  0.26  &  60  &  7205294  &  0.34  &  123  \\
1516253  &  0.13  &  68  &  4000693  &  0.18  &  60  &  4120506  &  0.15  &  56  &  4334234  &  0.32  &  158  &  7039590  &  0.29  &  70  &  7205482  &  0.35  &  66  \\
1516254  &  0.40  &  136  &  4000844  &  0.34  &  94  &  4120507  &  0.17  &  56  &  4336064  &  0.23  &  27  &  7040072  &  0.32  &  178  &  7205717  &  0.37  &  284  \\
1516257  &  0.14  &  68  &  4000845  &  0.37  &  94  &  4120508  &  0.04  &  56  &  4336209  &  0.31  &  128  &  7041290  &  0.11  &  300  &  7206507  &  0.05  &  148  \\
1531935  &  0.02  &  264  &  4000846  &  0.31  &  94  &  4122121  &  0.24  &  252  &  4337020  &  0.23  &  130  &  7045255  &  0.11  &  84  &  7207873  &  0.25  &  464  \\
1532799  &  0.09  &  120  &  4000852  &  0.33  &  94  &  4125718  &  0.30  &  88  &  4337186  &  0.21  &  250  &  7046163  &  0.37  &  250  &  7207874  &  0.23  &  464  \\
1542739  &  0.35  &  24  &  4000853  &  0.34  &  94  &  4125782  &  0.10  &  92  &  4337453  &  0.22  &  396  &  7047531  &  0.12  &  252  &  7208625  &  0.36  &  80  \\
1546845  &  0.20  &  146  &  4000854  &  0.31  &  94  &  4301919  &  0.17  &  244  &  4337700  &  0.34  &  346  &  7047694  &  0.31  &  148  &  7208879  &  0.26  &  260  \\
1550485  &  0.19  &  144  &  4000867  &  0.35  &  212  &  4302820  &  0.33  &  78  &  4338837  &  0.20  &  262  &  7047891  &  0.31  &  328  &  7209252  &  0.23  &  304  \\
2003573  &  0.20  &  226  &  4000868  &  0.28  &  212  &  4303361  &  0.33  &  240  &  4338838  &  0.33  &  342  &  7048400  &  0.39  &  432  &  7209599  &  0.17  &  120  \\
2003618  &  0.36  &  96  &  4000869  &  0.16  &  212  &  4303426  &  0.22  &  68  &  4339549  &  0.07  &  228  &  7051187  &  0.14  &  104  &  7210187  &  0.34  &  236  \\
2004926  &  0.14  &  248  &  4001466  &  0.34  &  96  &  4303427  &  0.10  &  80  &  4339550  &  0.17  &  228  &  7052148  &  0.22  &  456  &  7210622  &  0.06  &  180  \\
2005525  &  0.14  &  332  &  4001756  &  0.23  &  117  &  4304364  &  0.31  &  100  &  4340584  &  0.25  &  62  &  7052160  &  0.36  &  116  &  7211348  &  0.38  &  68  \\
2006386  &  0.13  &  111  &  4027976  &  0.18  &  252  &  4304826  &  0.13  &  184  &  4341452  &  0.33  &  196  &  7052161  &  0.28  &  66  &  7212738  &  0.03  &  256  \\
2006447  &  0.10  &  76  &  4030446  &  0.06  &  76  &  4307236  &  0.12  &  148  &  4342967  &  0.29  &  172  &  7052500  &  0.35  &  112  &  7212881  &  0.24  &  200  \\
2006640  &  0.38  &  156  &  4030472  &  0.03  &  356  &  4307335  &  0.23  &  180  &  4344542  &  0.36  &  116  &  7053319  &  0.38  &  408  &  7214394  &  0.26  &  116  \\
2007377  &  0.04  &  152  &  4030948  &  0.28  &  132  &  4308271  &  0.20  &  388  &  4345710  &  0.08  &  208  &  7057131  &  0.06  &  178  &  7215651  &  0.03  &  276  \\
2009382  &  0.34  &  188  &  4033562  &  0.40  &  128  &  4308985  &  0.29  &  180  &  4500101  &  0.38  &  58  &  7057715  &  0.36  &  270  &  7215652  &  0.06  &  276  \\
2009771  &  0.23  &  196  &  4060158  &  0.24  &  500  &  4310018  &  0.02  &  224  &  4500256  &  0.38  &  444  &  7057751  &  0.22  &  88  &  7215919  &  0.09  &  348  \\
2009875  &  0.26  &  28  &  4060716  &  0.12  &  110  &  4310046  &  0.20  &  200  &  4500879  &  0.36  &  107  &  7100829  &  0.06  &  76  &  7216865  &  0.22  &  80  \\
2010394  &  0.25  &  100  &  4076655  &  0.09  &  178  &  4310802  &  0.32  &  164  &  4501229  &  0.35  &  64  &  7101802  &  0.19  &  432  &  7217861  &  0.25  &  404  \\
2012185  &  0.20  &  168  &  4076797  &  0.33  &  500  &  4311621  &  0.40  &  45  &  4505752  &  0.30  &  54  &  7103562  &  0.24  &  392  &  7218742  &  0.36  &  30  \\
2012571  &  0.18  &  108  &  4076838  &  0.14  &  444  &  4311623  &  0.37  &  138  &  4506029  &  0.32  &  180  &  7103978  &  0.31  &  327  &  7218743  &  0.06  &  60  \\
2012640  &  0.02  &  82  &  4076839  &  0.37  &  456  &  4311749  &  0.27  &  100  &  4506561  &  0.13  &  160  &  7105225  &  0.36  &  116  &  7219263  &  0.18  &  78  \\
2015898  &  0.07  &  220  &  4077085  &  0.22  &  436  &  4311792  &  0.35  &  142  &  4506562  &  0.24  &  60  &  7106265  &  0.22  &  96  &  7219393  &  0.34  &  208  \\
2016308  &  0.25  &  192  &  4077087  &  0.33  &  500  &  4312085  &  0.20  &  88  &  4506830  &  0.11  &  68  &  7108362  &  0.21  &  250  &  7220737  &  0.12  &  138  \\
2017123  &  0.06  &  160  &  4080787  &  0.15  &  224  &  4312086  &  0.29  &  156  &  4506832  &  0.25  &  68  &  7110016  &  0.13  &  432  &  7223625  &  0.38  &  82  \\
2017434  &  0.17  &  204  &  4081130  &  0.34  &  268  &  4312745  &  0.01  &  120  &  4510004  &  0.35  &  52  &  7111061  &  0.20  &  356  &  7224170  &  0.27  &  204  \\
2019435  &  0.10  &  88  &  4100785  &  0.37  &  152  &  4313081  &  0.38  &  64  &  4510164  &  0.33  &  80  &  7111449  &  0.30  &  216  &  7224812  &  0.32  &  276  \\
2020191  &  0.39  &  104  &  4100786  &  0.37  &  74  &  4313572  &  0.05  &  136  &  7002466  &  0.03  &  496  &  7111663  &  0.37  &  188  &  7224813  &  0.08  &  92  \\
2101762  &  0.23  &  168  &  4101559  &  0.27  &  216  &  4313682  &  0.33  &  236  &  7005654  &  0.06  &  88  &  7111840  &  0.34  &  312  &  7225192  &  0.35  &  212  \\
2101805  &  0.16  &  352  &  4101560  &  0.16  &  108  &  4314012  &  0.28  &  112  &  7006006  &  0.10  &  174  &  7111918  &  0.24  &  88  &  7226282  &  0.26  &  95  \\
2103415  &  0.07  &  96  &  4101561  &  0.32  &  108  &  4314265  &  0.06  &  262  &  7007965  &  0.32  &  492  &  7114058  &  0.15  &  422  &  7226283  &  0.39  &  368  \\
2103416  &  0.07  &  108  &  4102190  &  0.23  &  68  &  4314268  &  0.02  &  262  &  7008910  &  0.32  &  280  &  7114953  &  0.18  &  168  &  7226286  &  0.35  &  408  \\
2104548  &  0.09  &  21  &  4102191  &  0.10  &  68  &  4314270  &  0.12  &  262  &  7010128  &  0.29  &  226  &  7115199  &  0.03  &  312  &  7226287  &  0.13  &  236  \\
2104553  &  0.14  &  38  &  4102192  &  0.28  &  68  &  4314271  &  0.23  &  262  &  7010621  &  0.31  &  150  &  7116032  &  0.33  &  440  &  7227238  &  0.28  &  128  \\
2104589  &  0.33  &  92  &  4102656  &  0.33  &  38  &  4315110  &  0.25  &  448  &  7011125  &  0.10  &  300  &  7122604  &  0.31  &  284  &  7228244  &  0.34  &  110  \\
2105778  &  0.25  &  60  &  4103043  &  0.03  &  404  &  4315143  &  0.10  &  220  &  7011126  &  0.14  &  150  &  7123550  &  0.15  &  302  &  7228672  &  0.35  &  44  \\
2108137  &  0.37  &  41  &  4104399  &  0.22  &  68  &  4316167  &  0.25  &  52  &  7011127  &  0.39  &  162  &  7151581  &  0.25  &  36  &  7229554  &  0.02  &  356  \\
2108147  &  0.38  &  112  &  4104401  &  0.08  &  68  &  4316171  &  0.33  &  25  &  7012773  &  0.38  &  102  &  7154787  &  0.27  &  180  &  7230473  &  0.28  &  158  \\
2108148  &  0.33  &  112  &  4104405  &  0.23  &  43  &  4316834  &  0.35  &  146  &  7012775  &  0.38  &  62  &  7200911  &  0.18  &  76  &  7230813  &  0.05  &  346  \\
2202012  &  0.14  &  52  &  4104409  &  0.20  &  43  &  4316896  &  0.29  &  120  &  7014286  &  0.32  &  308  &  7200913  &  0.08  &  76  &  7230814  &  0.14  &  334  \\
2203690  &  0.26  &  156  &  4104410  &  0.21  &  43  &  4318081  &  0.10  &  50  &  7014287  &  0.32  &  158  &  7200914  &  0.07  &  76  &  7230817  &  0.25  &  352  \\
2206371  &  0.38  &  120  &  4105695  &  0.40  &  500  &  4318432  &  0.28  &  57  &  7014288  &  0.32  &  304  &  7201051  &  0.03  &  276  &  7231268  &  0.11  &  256  \\
2207901  &  0.22  &  112  &  4106175  &  0.02  &  364  &  4322024  &  0.38  &  244  &  7014292  &  0.03  &  154  &  7201052  &  0.06  &  276  &  7232062  &  0.19  &  344  \\
2208031  &  0.36  &  76  &  4106572  &  0.34  &  250  &  4323190  &  0.24  &  320  &  7014446  &  0.30  &  84  &  7201329  &  0.09  &  58  &  7232249  &  0.18  &  100  \\
2208799  &  0.29  &  108  &  4109852  &  0.25  &  104  &  4323773  &  0.34  &  95  &  7014583  &  0.11  &  76  &  7201939  &  0.09  &  312  &  7232828  &  0.03  &  128  \\
2209735  &  0.37  &  140  &  4110042  &  0.12  &  122  &  4323810  &  0.04  &  188  &  7014584  &  0.14  &  76  &  7202467  &  0.18  &  52  &  8000064  &  0.14  &  84  \\
2209749  &  0.07  &  200  &  4110390  &  0.39  &  142  &  4324027  &  0.17  &  288  &  7015536  &  0.13  &  98  &  7202481  &  0.35  &  228  &  8100077  &  0.15  &  108  \\
2211667  &  0.19  &  150  &  4110455  &  0.24  &  252  &  4325015  &  0.40  &  43  &  7015537  &  0.36  &  98  &  7202485  &  0.32  &  114  &  8100483  &  0.09  &  148  \\
2213373  &  0.14  &  320  &  4111467  &  0.39  &  236  &  4325471  &  0.28  &  62  &  7015915  &  0.33  &  296  &  7203118  &  0.37  &  164  &  8102065  &  0.38  &  204  \\
2214075  &  0.09  &  50  &  4112026  &  0.35  &  228  &  4327550  &  0.40  &  248  &  7015917  &  0.30  &  156  &  7203121  &  0.38  &  184  &    &    &    \\
    \end{tabular}
\end{table}
\end{widetext}

\end{document}